\documentstyle[floats,epsf,prbbib,aps]{revtex}

\begin{document}

\twocolumn[

\title{\vspace*{-1cm}\hfill 
{\tt Submitted to Phys.~Rev.~B}
       \vspace{1cm}\\
The GaAs Equilibrium Crystal Shape from First-Principles}

\author{N.~Moll\cite{Author}, A.~Kley, E.~Pehlke, and M.~Scheffler}
\address{
Fritz-Haber-Institut der Max-Planck-Gesellschaft, Faradayweg 4-6,
D-14195 Berlin-Dahlem, Germany}
\date{\today}
\maketitle

\begin{abstract}
Surface energies for different GaAs surface orientations have been
calculated as a function of the chemical potential. We use an energy
density formalism within the first-principles pseudopotential
density-functional approach. The equilibrium crystal shape (ECS) has
been derived from the surface energies for the (110), (100), (111),
and (\=1\=1\=1) orientations. Under As-rich conditions all four
considered surface orientations exist in thermodynamic equilibrium, in
agreement with experimental observations. Moreover, our calculations
allow us to decide on previous contradictory theoretical
values for the surface energies of the (111) and (\=1\=1\=1) facets.
\end{abstract}

% meaning of PACS numbers:
% 68.35.Bs Surface structure and topography
% 68.35.Md Surface energy; thermodynamic properties (see also 82.65.D
%          Thermodynamics of surfaces and interfaces)
% 73.20.-r Surface and interface electron states
]

\thispagestyle{empty}

\section{Introduction}

The equilibrium crystal shape (ECS) is that shape which, in the limit
of infinitely large volume, yields the minimum free energy of a
crystal.  For a given arbitrary surface orientation and unit cell the
atomic reconstruction that yields the lowest surface free energy can
be determined.  However, it is well known that in general this will
not result in a thermodynamically stable situation, because the
surface can further lower its energy by faceting on a macroscopic
scale. The ECS provides a set of surface orientations that exist in
thermodynamic equilibrium. Except for some situations with degenerate
surface energies surfaces of any other orientations will facet.

The faceting of GaAs surfaces has been studied experimentally. Whereas
Weiss {\em et al}.\ \cite{weiss:89} studied the different surface
orientations exposed on a round shaped crystal with low-energy
electron-diffraction (LEED), N\"otzel {\em et al}.\ \cite{noetzel:92}
investigated various planar high-index surfaces with reflection
high-energy electron-diffraction (RHEED). Both groups observed for
different high-index surface orientations faceting into low-index
surfaces. Moreover, surface energies play a major role in the
formation of islands during heteroepitaxy. For example, InAs grows on
GaAs in the Stranski-Krastanov mode. \cite{leonard:93} The surface
energy of InAs being lower than that of GaAs, first a uniform
wetting-layer forms.  During further deposition of InAs
three-dimensional islands are formed due to strain relaxation.
Recently, these quantum dots have attracted great
interest. \cite{leonard:93,moison:94,grundmann:95,shchukin:95} Besides
other quantities like the elastic relaxation energy of the islands,
the absolute InAs surface energies of the involved facets which we
assume to be similar to those of GaAs enter into the theory of the
shape and size of the islands.

Both experimental as well as calculated {\em absolute} values of the
surface energy as a function of orientation are quite scarce.  The
surface energy has been measured for the GaAs (110) surface in a
fracture experiment.\cite{messmer:81} Relative surface energies and
the ECS of Si have been determined \cite{eaglesham:93,bermond:95}, but
to our knowledge no such measurements have been carried through for
GaAs.  Moreover, it is often difficult to establish whether an
observed surface really represents thermodynamic equilibrium.  At low
temperatures faceting and therefore thermodynamic equilibration may be
hindered by insufficient material transport. At high temperatures,
kinetics may govern the surface morphologies due to evaporation.

The purpose of this work is to present the {\em absolute} values for
the surface energy of the GaAs (110), (100), (111), and (\=1\=1\=1)
surfaces calculated from first principles, and the ECS constructed
from these data. Empirical potentials do not produce reliable surface
properties. {\em Ab initio} calculations have been carried out by
various groups for different surface orientations of GaAs.  Qian {\em
et al}.\ \cite{qian:88a} used an {\em ab initio} pseudopotential
method to calculate the absolute surface energy of the GaAs (110)
surface. They found very good agreement with the experimental cleavage
energy.  Northrup and Froyen \cite{northrup:93}, Qian {\em et al}.\
\cite{qian:88}, and Ohno \cite{ohno:93} determined the (100)
reconstruction with lowest energy. The absolute surface energies for
these reconstructions were not given, however. Kaxiras {\em et al}.\
\cite{kaxiras:87} calculated energies for GaAs (111) reconstructions
relative to the surface energy of the ideal (111) surface. For the
(\=1\=1\=1) surface Kaxiras {\em et al}.\ \cite{kaxiras:87a} and
Northrup {\em et al.}\ \cite{biegelsen:90} calculated relative surface
energies for different $(2 \times 2)$ reconstructions. Based on their
results they predicted the (\=1\=1\=1) equilibrium reconstruction.

However, for geometrical reasons it is impossible to derive absolute
surface energies for the (111) and (\=1\=1\=1) orientations of GaAs
from such total-energy calculations. Chetty and Martin
\cite{chetty:92,chetty:92a} solved this problem by introducing an
energy density, which enables the computation of the energies of the
top and the bottom surfaces of the slab separately.  Having calculated
the absolute surface energies for the ideal reference surfaces they
transformed the relative surface energies of Kaxiras {\em et al}.\
\cite{kaxiras:87,kaxiras:87a} and Northrup {\em et al.}\
\cite{biegelsen:90} to absolute surface energies. A comparison of
these absolute values, however, shows that the two results differ
significantly. This difference is not yet understood, and we will come
back to it in Section IV below.

We have calculated absolute surface energies for the different
orientations directly (i.e., without introducing a reference surface)
and consistently with one and the same set of parameters and
pseudo-potentials. Before we will detail our results and the ECS of
GaAs in Section IV, we will first give an overview of GaAs surface
properties in Section II and describe the computational details in
Section III.

\section{Chemical Potential and Surface Reconstruction}
 
The stable surface reconstruction is the one with the lowest 
surface free energy. In our case the substrate consists of two 
elements and thus the difference of the number of atoms of the 
two species enters as another degree of freedom in addition to 
the atomic geometry. Non-stoichiometric surfaces are considered
by allowing the surface to exchange atoms
with a reservoir, which is characterized by a chemical potential.
The equilibrium is determined by the minimum of the free energy
\begin{equation}
  \gamma_{\rm surface} A = E_{\rm surface} - \sum_i \mu_i N_i .
\end{equation}
The surface free energy $\gamma_{\rm surface}A$ of the surface area
$A$ has been calculated for zero temperature and pressure and
neglecting zero point vibrations. The chemical potential $\mu_i$ is
the free energy per particle in the reservoir for the species $i$, and
$N_i$ denotes the number of particles of the species $i$. The
temperature dependence is ignored because the contributions tend to
cancel for free energy differences.

In experiment the value of the chemical potential can be varied over a
certain interval. This interval is limited by the bulk chemical
potentials of the condensed phases of Ga and As
\cite{qian:88,biegelsen:90}, corresponding to the two following
situations: On the one hand the surface can be in equilibrium with
excess Ga-metal, which has the chemical potential $\mu_{\rm Ga
(bulk)}$, and the GaAs bulk with chemical potential $\mu_{\rm GaAs}$.
On the other hand the surface can be in equilibrium with bulk As and,
again, the GaAs bulk. Both reservoirs can act as sinks and sources of
surface atoms. The upper limit of each chemical potential is
determined by the condensed phase of the respective element,
\begin{equation}
  \mu_i < \mu_{i \rm (bulk)},
  \label{muup}
\end{equation}
because otherwise the elemental phase would form on the GaAs 
surface. Furthermore, in thermodynamic equilibrium the sum of chemical 
potentials of Ga and As must be equal to the bulk energy per GaAs pair,
\begin{eqnarray}
 \mu_{\rm Ga} + \mu_{\rm As} & = & \mu_{\rm GaAs} \nonumber\\
  & = & \mu_{\rm Ga (bulk)} + \mu_{\rm As (bulk)} - \Delta H_f.
  \label{heat}
\end{eqnarray}
For the heat of formation we have calculated a value of 0.64 eV using
a plane-wave cutoff of 10 Ry which is in good agreement with the
experimental value \cite{hcp:86} of 0.74 eV. For the bulk calculations
we computed the bulk energy of Ga in an orthorhombic structure
\cite{qian:88} and the bulk energy of As in a trigonal structure
\cite{needs:86}.
 
In this work we give the surface energies in dependence of the As
chemical potential. Therefore, we write equations (\ref{muup}) and
(\ref{heat}) in the following form
\begin{equation}
  \mu_{\rm As (bulk)} - \Delta H_f < \mu_{\rm As} < \mu_{\rm As (bulk)}.
\end{equation}
The surface energy is calculated from the total energy $E_{\rm tot}$, 
\begin{equation}
  \gamma_{\rm surface} A = E_{\rm tot} - \mu_{\rm GaAs} N_{\rm Ga}
  - \mu_{\rm As}(N_{\rm As} - N_{\rm Ga}).
  \label{esurface}
\end{equation}
The stoichiometry of the surface, $\Delta N = N_{\rm As} - N_{\rm
Ga}$, determines the slope of the surface energy versus the chemical
potential.  A consistent counting method for $\Delta N$ has to be
applied to all orientations. We apply the method of Chetty and Martin
\cite{chetty:91} which utilizes the bulk symmetries of the
crystal. For example, following their counting method the ideal (110)
cleavage surface is stoichiometric, i.e.\ the difference $\Delta N$ is
equal to zero. Thus the surface energy of the (110) cleavage surface
is independent of the chemical potential.
 
\begin{figure}[tb]
  \epsfxsize=8.7cm
  \epsfbox{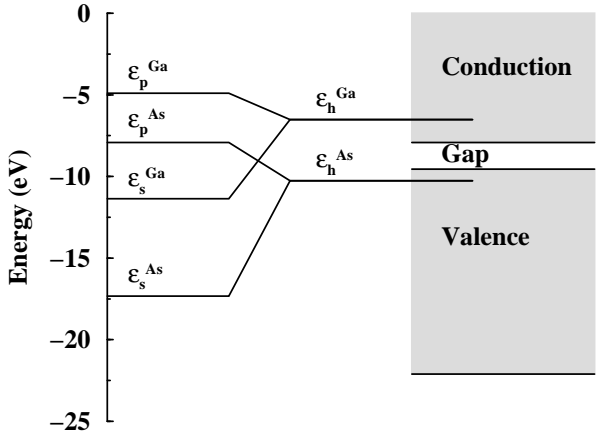}
  \caption{The energy levels of the $s$ and $p$ orbitals 
    $\epsilon_{s,p}$, of the $sp^3$ dangling-bonds $\epsilon_{\rm h}$,
    and of the conduction and valence band. The data are from Harrison.
    \protect\cite{harrison:79}}
  \label{energy_levels}
\end{figure}

When the chemical potential is varied, different reconstructions with
different surface stoichiometries become thermodynamically stable.
All experimentally observed reconstructions, however, fulfill certain
conditions. First of all, GaAs surfaces favor to be semiconducting, as
this leads to a low surface energy. Surface bands in the bulk gap and
especially surface bands crossing the Fermi-level will lead to a
higher surface energy. The electron counting model
\cite{harrison:79,pashley:89} gives a simple criterion whether a
surface can be semiconducting or not.  In the bulk the $sp^3$
hybridized orbitals of GaAs form bonding and antibonding states. At
the surface there are partially filled dangling bonds. Their energies
are shown schematically in Fig.\ \ref{energy_levels}, they are
estimated from the atomic $s$ and $p$ eigenenergies of either species.
Compared to the dispersion of the conduction and the valence bands,
the dangling bond energy of the cation (Ga) falls into the conduction
band and therefore it should be empty. The dangling bond energy of the
anion (As) lies in the valence band and therefore it should be
filled.  Thus there has to occur an electron transfer from the Ga to
the As dangling bonds.  For a low-energy semiconducting surface the
dangling bonds in the conduction band have to be empty, exactly
filling all the dangling bonds in the valence band. Otherwise the
surface becomes metallic and has a higher surface energy.  Ga and As
surface atoms are added to, or removed from, the ideal bulk-truncated
polar surfaces to obtain a low-energy semiconducting surface.
 
Secondly, the electron transfer from the Ga dangling bonds to the As
dangling bonds has consequences for the geometry of the surface
reconstructions. The surface Ga atom which has lost an electron favors
a $sp^2$ like hybridization. Therefore the Ga atom relaxes inwards and
forms a more planar configuration. The dangling bond of arsenic is
completely filled and the As atom prefers to form bonds with its three
$p$ orbitals. Therefore the bond angle of the surface As atom is close
to 90$^\circ$, and the As atom relaxes outwards. These configurations
resemble the bond geometry of small molecules like GaH$_3$ and AsH$_3$
and are a general result for surfaces of III-V
semiconductors. \cite{alves:91}

\section{Computational Details}

To determine the surface energies we carried out total-energy
calculations using density-functional
theory.\cite{hohenberg:64,kohn:65} We applied the local-density
approximation to the exchange-correlation functional, choosing the
parameterization by Perdew and Zunger \cite{perdew:81} of Ceperley and
Alder's \cite{ceperley:80} data for the correlation energy of the
homogeneous electron gas. The surfaces were described by periodically
repeated slabs. All computations were done with an extended version of
the computer code {\em fhi93cp}. \cite{stumpf:94} The program employs
{\em ab initio} pseudopotentials and a plane-wave basis-set. It was
generalized to additionally compute the energy density according to
Chetty and Martin. \cite{chetty:92}

The slab geometry leads to serious problems when surface energies of
zinc-blende structures are to be calculated for arbitrary
orientations.  To derive the surface energy from a total energy
calculation both surfaces of the slab have to be equivalent.  Though
such slabs can be constructed for the (110) and the (100) orientation,
this is impossible for the (111) orientation: The (111) and the
(\=1\=1\=1) surfaces of GaAs are inequivalent. This follows from the
simple geometric property of the zinc-blende structure that the Ga-As
double layers are Ga and As terminated on the top and bottom side of
the slab, respectively.  Chetty and Martin\cite{chetty:92} solved this
problem by introducing an energy density.  The energy density itself,
however, does not bear any physical significance, only the integrals
of the energy density over suitable parts of the supercell (e.g.,
volumes bounded by bulk mirror planes) lead to well-defined,
physically meaningful energies. \cite{chetty:92} We have checked the
accuracy of this approach for our GaAs slabs: Variation of the surface
reconstruction on the bottom side of the (100) and (111) slabs results
in a negligible change of the surface energy of the surface on the top
($<0.7$ meV/{\AA}$^2$).

{\em Ab initio} norm-conserving pseudopotentials were generated with
Hamann's scheme. \cite{hamann:89} The cutoff radii for pseudoization
have been chosen equal to 0.58 {\AA}, 0.77 {\AA}, and 1.16 {\AA} for
the $s$, $p$, and $d$ wave-functions of Ga, and equal to 0.61 {\AA},
0.60 {\AA}, and 1.07 {\AA} for $s$, $p$, and $d$ wave-functions of
As. The semi-local pseudopotentials were further transformed into
fully separable Kleinman-Bylander pseudopotentials \cite{kleinman:82},
with the $d$ potential chosen as the local potential. The logarithmic
derivatives of the different potentials were examined and various
transferability tests \cite{stumpf:91}, e.g.\ ``hardness'' tests, were
performed. All together the potentials showed good transferability.
The structures of the bulk phases of Ga and As are well described by
these potentials, the theoretical lattice constants being only
slightly smaller than the experimental ones with a relative deviation
below 3.5\%.

The wave functions were expanded into plane waves \cite{ihm:79} with a
kinetic energy up to 10 Ry. This leads to a convergence error in the
surface energies of less than 3 meV/{\AA}$^2$. The electron density
was calculated from special {\bf k}-point sets \cite{monkhorst:76},
their density in reciprocal space being equivalent to 64 {\bf
k}-points in the whole (100) ($1 \times 1$) surface Brillouin-zone.

For the (100), (111), and (\=1\=1\=1) surfaces ``pseudo-hydrogen'' was
used to saturate the bottom surfaces of the slabs.
\cite{shiraishi:90} Pseudo-hydrogen denotes a Coulomb-potential with a
non-integer core-charge $Z$, together with $Z$ electrons. The Ga and
As atoms of these surfaces were fixed at their ideal bulk
positions. The Ga terminated surface was saturated with
pseudo-hydrogen with an atomic number of $Z=1.25$. On each dangling
bond of a Ga surface atom one pseudo-hydrogen was placed. Similarly,
the As terminated surface was saturated with pseudo-hydrogen with an
atomic number of 0.75. The saturated surfaces are semiconducting
without any surface states in the bulk band-gap. There are two main
advantages using this pseudo-hydrogen. First of all, the interaction
of both surfaces with each other is in this way minimal. Secondly, the
surface atoms which are saturated with the pseudo-hydrogen can be kept
fixed at ideal bulk positions. Thus thinner slabs can be used and
charge sloshing is suppressed.

For polar surfaces, such as the ideal (111) surface, a difficulty
arises due to charge transfer from one side of the slab to the
opposite side. This charge transfer is hindered by a semiconducting
surface, e.g. the pseudo-hydrogen saturated surface at the bottom of
the slab. We estimate the uncertainty due to charge transfer to be
smaller than 1.4 meV/{\AA}$^2$ for a polar surface, comparing the
surface energies of the pseudo-hydrogen saturated surface derived from
two calculations. One is carried through with a semiconducting surface
on the top of the slab, the other one with a metallic surface.

We have carried out computations for a large variety of
reconstructions of the GaAs (110), (100), (111), and (\=1\=1\=1)
surfaces, which have previously been suggested in literature. Starting
from some initial geometry, the atom positions in the topmost layers
of the slab were relaxed until the forces on the atoms were smaller
than 50 meV/{\AA}.  The other layers were kept fixed at their ideal
bulk positions with a bulk lattice-constant of 5.56 \AA which had been
determined theoretically at the same cutoff energy as the slab
calculations and using 384 {\bf k}-points in the whole
Brillouin-zone. This value is 1.4\% smaller than the experimental
lattice constant\cite{landolt:82} neglecting zero point vibrations.
 
\section{Results and Discussion}
\subsection{(110) Surface}

The (110) surface is one of the most extensively studied GaAs surfaces
(Ref.\ \onlinecite{qian:88a,kuebler:80,northrup:91,alves:91} and
references therein). The (110) plane is the cleavage plane of III-V
semiconductors. Containing the same number of cations (Ga) and anions
(As) it is intrinsically neutral. The cleavage surface does not
reconstruct, only a relaxation of surface atomic positions within the
$(1 \times 1)$ surface unit cell is observed. The charge from the Ga
dangling-bond is transfered into the As dangling-bond, which becomes
completely filled. The orbitals of both surface atoms rehybridize, and
the zigzag chains of Ga and As surface atoms tilt, with the As atom
being raised and the Ga atom being lowered. Thereby the Ga surface
atom acquires a nearly planar bonding configuration, while the As
surface atom relaxes towards a pyramidal configuration with orthogonal
bonds.

\begin{figure}[tb]
  \epsfxsize=7.8cm
  \hspace*{0.45cm}\epsfbox{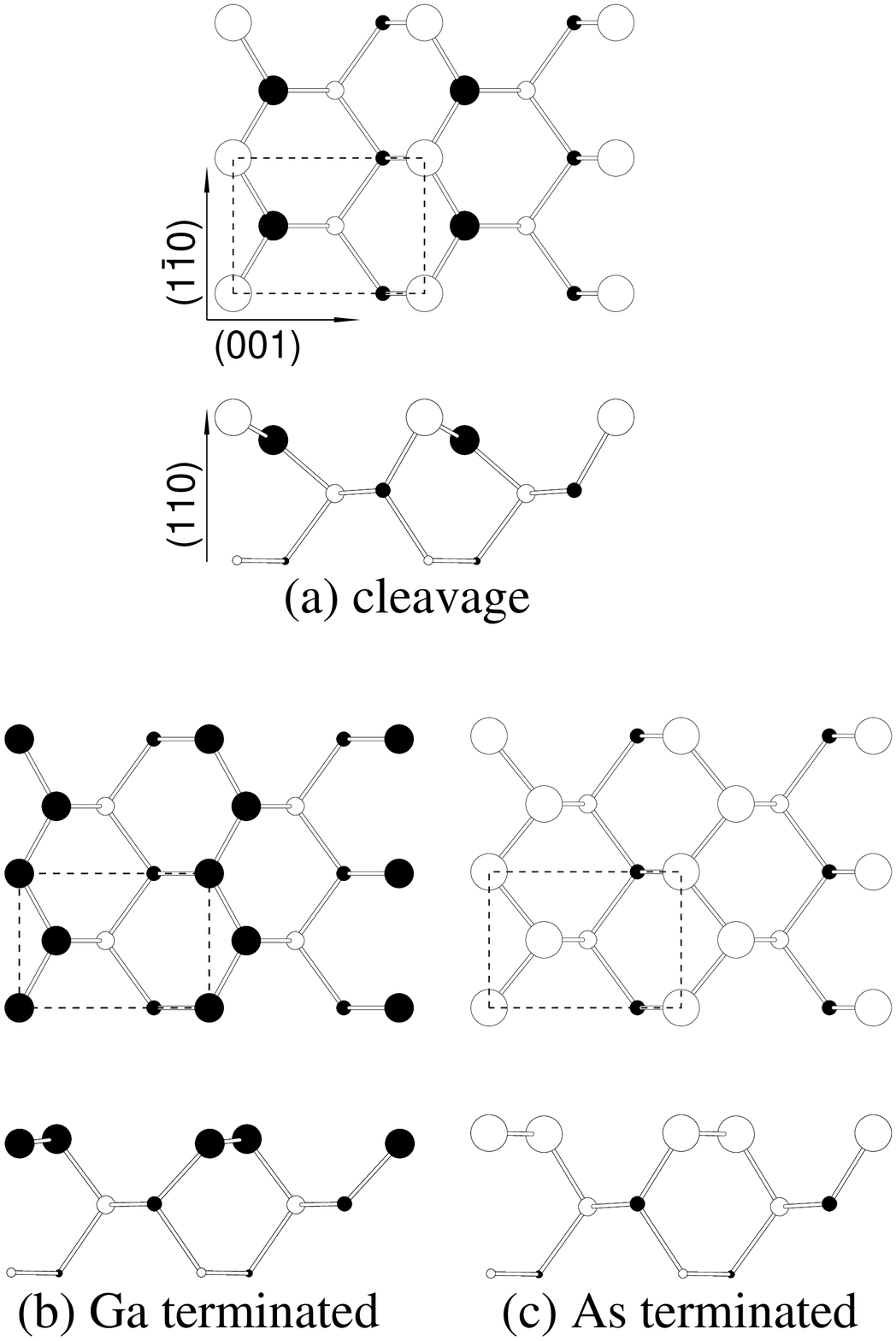}
  \caption{Atomic structures of the GaAs (110) surface in
    top and side view. Open and filled circles denote As and Ga atoms, 
    respectively.}
  \label{110_geo}
\end{figure}

We have calculated the surface energy of the relaxed cleavage surface
shown in Fig.\ \ref{110_geo}(a). It is stoichiometric ($\Delta N = 0$)
and semiconducting.
In addition, we considered two other surface structures: The Ga
terminated (110) surface is shown in Fig.\ \ref{110_geo}(b).  Formally
it can be constructed from the cleavage surface by substituting all
top-layer As atoms by Ga atoms. This surface has a stoichiometry of
$\Delta N = -2$ per $(1 \times 1)$ cell, and it fulfills the electron
counting criterion. Nevertheless, it is not semiconducting, because
the bands of the Ga-Ga surface bonds cross the Fermi-level. The Ga
surface atoms do not relax in the same way as the respective Ga and As
atoms in the cleavage surface, instead they almost stay in the same
plane. Finally, we have calculated the surface energy of the As
terminated (110) surface (see Fig.\ \ref{110_geo}(c)). Here the Ga
surface atoms have been replaced by As atoms, which yields a surface
with a stoichiometry of $\Delta N = 2$ per $(1 \times 1)$ surface unit
cell. Also this surface fulfills the electron counting criterion, and
it is semiconducting. Both As dangling-bonds are completely filled and
lie beneath the Fermi-level.  Similar to the Ga terminated surface,
the As surface atoms do not relax significantly, but stay in the same
plane.

For all three (110) surface reconstructions we used the same super
cell, with slabs composed of nine atomic layers and a vacuum region
with a thickness equivalent to seven atomic layers. The whole surface
Brillouin zone was sampled with 48 special {\bf k}-points. 
\cite{monkhorst:76}

\begin{figure}[tb]
  \epsfxsize=8.7cm
  \epsfbox{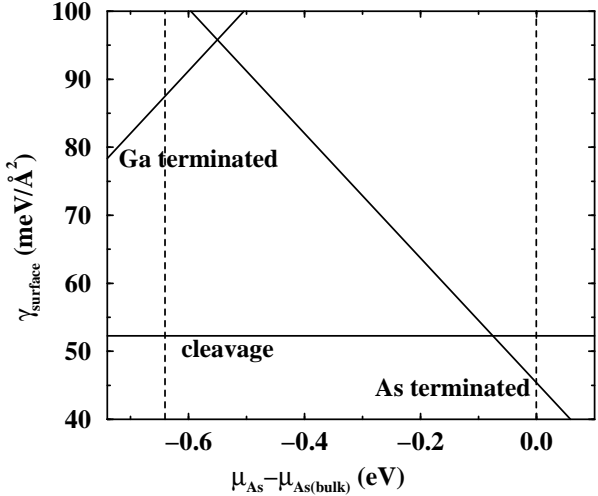}
  \caption{Surface energy of the different GaAs (110)
    surface structures in meV/{\AA}$^2$ plotted versus the 
    difference of the chemical potential of As and As bulk.}
  \label{110}
\end{figure}

The calculated surface energies are shown in Fig.\ \ref{110} for the 
three surface structures we have considered.
For a large range of the chemical potential the cleavage surface is
energetically most favorable. Our result for the surface energy of 52
meV/{\AA}$^2$ is in good agreement with the value of 57eV/{\AA}$^2$
meV which was calculated by Qian {\em et al}.\ \cite{qian:88a} using
essentially the same {\em ab initio} method. Both results compare very
well with the experimental surface energy of 54 $ \pm $ 9
meV/{\AA}$^2$ from fracture experiments by Messmer and Bilello.
\cite{messmer:81} In As-rich environments we find the As terminated
surface to exist in thermodynamical equilibrium, in agreement with
Northrup's calculation.\cite{northrup:91} We obtain a value of 45
meV/{\AA}$^2$ for the surface energy in an As-rich environment.
K\"ubler {\em et al.} \cite{kuebler:80} provided experimental evidence
for the existence of this structure. Using LEED they observed that the
surface relaxation was removed as the As coverage was increased. In
contrast to the As terminated surface, we find the Ga terminated
surface to be unstable even under the most extreme Ga rich conditions.

\subsection{(100) Surface}

Among the different orientations the (100) surface is the one used
most widely for the growth of opto-electronic devices.  The (100)
surface is polar, i.e.\ the planes parallel to the surface consist of
either only Ga or only As atoms. As a consequence, the stable surface
structure \cite{pashley:89} displays various reconstructions which
distinctly differ from those found on the (100) faces of the covalent
group IV semiconductors. D\"aweritz {\em et al}.\ \cite{daeweritz:90}
have derived a steady state ``phase'' diagram for the surface
reconstruction as a function of growth conditions. In their diagram
they point out 14 different reconstructions. To our knowledge, the
{\em equilibrium} phase diagram of the (100) surface has not yet been
determined. However, there are certain reconstructions which are
generally observed during and after growth. While heating the surface
Biegelsen {\em et al}.\ \cite{biegelsen:90a} observed a sequence of
phases from the As-rich $c(4 \times 4)$, $(2 \times 4)$ to the Ga-rich
$(4 \times 2)$ reconstructions. For each of these surface unit-cells
there exists a large variety of possible atomic configurations.

Chadi \cite{chadi:87} performed tight binding based total energy
minimizations to examine the structure of the $(2 \times 1)$ and $(2
\times 4)$ reconstructed surface. For the $(2 \times 4)$ he suggested
two possible atomic configurations with three and two As-dimers
($\beta$ and $\beta2$, notation according to Northrup {\em et al}.\
\cite{northrup:94}) per surface unit-cell. Moreover, he determined the
energy difference between the $(2 \times 4)$ and the related $c(2
\times 8)$ reconstruction to be less than 1 meV/{\AA}$^2$. As the $(2
\times 4)$ and the $c(2 \times 8)$ are very similar and have only
small difference in surface energy, we have not calculated the
centered reconstructions $c(2 \times 8)$ and $c(8 \times 2)$.
Ohno\cite{ohno:93} and Northrup\cite{northrup:93} carried through {\em
ab initio} calculations of the surface energies. Ohno could exclude
various configurations of the $(2 \times 1)$ and $(3 \times 1)$
surface unit-cell from being equilibrium structures. Moreover, he
concluded that for the $(2 \times 4)$ reconstruction the phase $\beta$
with three surface dimers is stable, which appeared to be in agreement
with the STM observations of Biegelsen {\em et
al}.\cite{biegelsen:90a} However, calculations by Northrup {\em et
al.}\ \cite{northrup:94} showed that the most stable $(2 \times 4)$
reconstruction contains two As-dimers in the top layer, which has been
confirmed by recent high resolution STM
observations. \cite{hashizume:94} Northrup {\em et al}.\ also
investigated the energetics of the $(4 \times 2)$ and $c(4 \times 4)$
reconstructions. For the $(4 \times 2)$ reconstruction they found a
two-dimer phase to be energetically favorable in agreement with STM
investigations.\cite{xue:95} However, a recent analysis of LEED
intensities by Cerd\a'a {\it et al.}\cite{cerda:95} suggests that the
top layer consist of three Ga dimers per $(4 \times 2)$ unit cell. For
the $c(4 \times 4)$ reconstruction Northrup {\em et al}.\ considered a
three-dimer phase\cite{biegelsen:90a} which they found to be stable in
certain conditions with respect to the $(2 \times 4)$ and $(4 \times
2)$ reconstructions. On the other hand a two-dimer phase was suggested
by Sauvage-Simkin {\em et al}.\ \cite{simkin:89} on the basis of X-ray
scattering experiments, and by Larsen {\em et al}.\ \cite{larsen:83}
who studied the surface with a number of different experimental
techniques.

In our calculations we have considered all atomic configurations with
a $(2 \times 4)$ and a $(4 \times 2)$ surface unit-cell that were
previously investigated by Northrup {\em et al}.
\cite{northrup:93,northrup:94} For the $c(4 \times 4)$ reconstruction
we took into account the three-dimer phase \cite{biegelsen:90a} and a
structure which has two instead of three As-dimers in the top
layer. \cite{simkin:89,larsen:83} The total energy calculations were
performed using supercells containing seven layers of GaAs. The
thickness of the vacuum corresponded to five layers GaAs.

\begin{figure}[tb]
  \epsfxsize=8.7cm
  \epsfbox{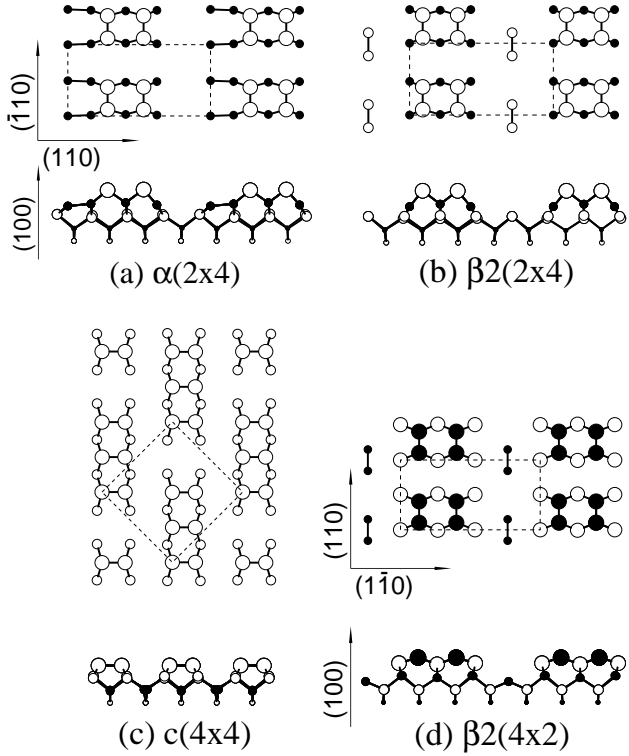}
  \caption{Atomic structures of the GaAs (100) reconstructions.}
  \label{100_geo}
\end{figure}

In Fig.\ \ref{100_geo} the geometries of those surface structures are
shown that have minimum surface energy within some range of the
chemical potential and therefore exist in thermodynamic equilibrium.
All four structures fulfill the electron counting criterion and are
semiconducting, i.e., the anion dangling bonds are filled and the
cation dangling bonds are empty. Furthermore the surfaces display
Ga-Ga bonds and As-As bonds, both having filled bonding and empty
antibonding states. The $\alpha(2 \times 4)$ reconstruction (Fig.\
\ref{100_geo}(a)) is stoichiometric ($\Delta N = 0 $). In the top
layer four As atoms are missing per $(2 \times 4)$ cell. The surface
As atoms form two dimers. The Ga-layer underneath is complete, but
differs from the bulk geometry by two Ga-Ga bonds which are formed
between the Ga atoms in the region of the missing As dimers. Removing
the Ga atoms in the missing dimer region one obtains the $\beta2(2
\times 4)$ structure in Fig.\ \ref{100_geo}(b) with a stoichiometry of
$\Delta N = \frac{1}{4}$ per $(1 \times 1)$ unit cell.  The completely
As-terminated $c(4 \times 4)$ surface shown in Fig.\ \ref{100_geo}(c)
has a stoichiometry of $\Delta N = \frac{5}{4}$ per $(1 \times 1)$
unit cell. It consists of three As-dimers which are bonded to a
complete As-layer beneath. The $\beta2(4 \times 2)$ structure shown in
Fig.\ \ref{100_geo}(d) represents the Ga-terminated counterpart of the
$\beta2(2 \times 4)$ reconstruction, with Ga atoms exchanged for As
atoms and vice versa. Thus the top layer consists of two Ga dimers per
$(4 \times 2)$ cell, and the second layer lacks two As atoms. This
results in a stoichiometry of $\Delta N = -\frac{1}{4}$ per 
$(1 \times 1)$ cell.

\begin{figure}[tb]
  \epsfxsize=8.7cm
  \epsfbox{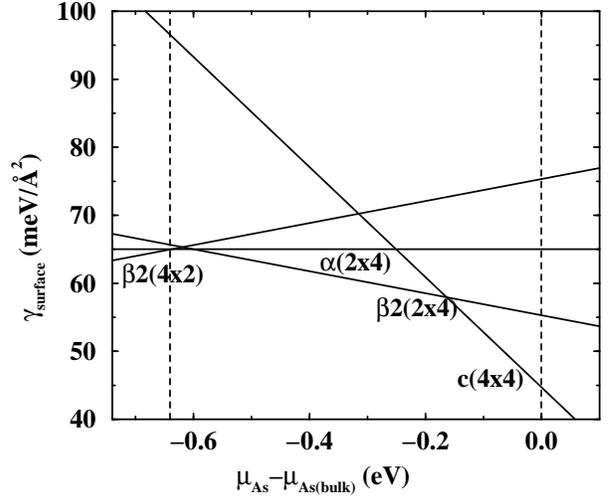}
  \caption{Surface energy of the different GaAs (100)
    reconstructions in meV/{\AA}$^2$ plotted versus the 
    difference of the chemical potential of As and As bulk.}
  \label{100}
\end{figure}

Our calculated surface energies of these four phases are
shown in Fig.\ \ref{100} as a function of the chemical potential.  
We predict the same sequence of equilibrium surface structures as
Northrup and Froyen \cite{northrup:93,northrup:94} as a function of
increasing As coverage: $\beta2$(4$\times$2), $\alpha$(2$\times$4),
$\beta2$(2$\times$4), and c(4$\times$4). The $c(4 \times 4)$ structure
with only two surface As-dimers per unit cell, which we considered in
addition to the structures investigated by Northrup and Froyen, turned
out to be unstable. Though this structure is more Ga-rich than the
c(4$\times$4) three As-dimer structure shown in Fig.\
\ref{100_geo}(c), even in the Ga-rich environment the two-dimer phase
has a surface energy which is 5 meV/{\AA}$^2$ higher than for the
three-dimer phase.  Due to the lack of absolute values in previous
calculations, quantitatively we can only compare energy differences
between surfaces with the same stoichiometry. Further comparison is
made difficult by the different range of the chemical potential in our
versus Northrup and Froyen's calculation
\cite{northrup:93,northrup:94}: Their value for the heat of formation
is $\Delta H_{f}$ = 0.92 eV, as opposed to our smaller value of
$\Delta H_{f}$ = 0.64 eV.  Comparing the three dimer phase $\beta$
with the two dimer phase $\beta2$, which both have the same
stoichiometry, we find that the two dimer phase has a surface energy
lower by 2 meV/{\AA}$^2$. This agrees with the result of Northrup and
Froyen who report an energy difference of 3 meV/{\AA}$^2$, and it
further confirms the conclusion that the three dimer phase $\beta$
does not exist in equilibrium.  On the whole, the agreement with the
relative surface energies calculated by Northrup {\em et al.} is
good. They can be converted to absolute surface energies by shifting
them by $\approx$ 65 meV/{\AA}$^2$, which results in a diagram similar
to Fig. \ref{100}.

All investigated (100) surfaces display similar atomic relaxations
which are characterized by the creation of dimers and the
rehybridization of threefold coordinated surface atoms.  The creation
of surface dimers decreases the number of partially occupied dangling
bonds, and by rehybridization the surface gains band structure
energy. The calculated bond lengths in bulk Ga and As, 2.32 {\AA} and
2.50 {\AA}, respectively, can serve as a first estimate for the
respective dimer bond lengths on the GaAs surface.  Our calculations
yield As-dimer lengths between 2.45 and 2.50 {\AA} for the $\alpha$
and $\beta2$ surface reconstruction. This is within the range of
experimentally deduced values which scatter between 2.2 and 2.9 {\AA}
\cite{chambers:92,xu:93,li:93} and it is similar to the dimer lengths
of 2.53 and 2.55 {\AA} which were determined by Northrup {\em et al}.
\cite{northrup:95} On the $c(4 \times 4)$ reconstructed surface the
calculated As-dimer lengths are 2.57 {\AA} for the central dimer and
2.53 {\AA} for the two outer dimers of the three-dimer strings in the
surface unit-cell. Using X-ray scattering Sauvage-Simkin {\em et al}.\
\cite{simkin:89} determined these bond-lengths as $2.63 \pm 0.06$
{\AA} and $2.59 \pm 0.06$ {\AA}. Very recently Xu {\it et al.}\
\cite{xu:95} suggested that the dimers on the c(4$\times$4) structure
should be tilted by $4.3^\circ$. However, as for the $(2 \times 4)$
reconstructions we find the dimers to be parallel to the surface, in
agreement with several previous
experiments\cite{biegelsen:90a,simkin:89}.  Even when starting with an
initial configuration with surface dimers tilted by $8^\circ$ we find
the dimers to relax back to the symmetric positions with a residual
tilt angle less than $0.1^\circ$. The Ga-Ga dimer bond length is
calculated to be 2.4 {\AA} on the $\beta2(4 \times 2)$ reconstruction
and 2.5 {\AA} on the $\alpha(2 \times 4)$ structure which agrees with
previous {\em ab initio} calculations. \cite{northrup:95} From a
recent LEED investigation of the Ga rich (100) surface Cerd\a'a {\em
et al}.\cite{cerda:95} deduced that the stable $(4 \times 2)$
reconstructed surface displays three Ga-dimers per unit cell with
unusual dimer lengths of 2.13 {\AA} and 3.45 {\AA}.  In our
calculation, however, this three dimer phase is energetically slightly
less favorable than the two dimer phase $\beta2(4 \times 2)$ by 0.8
meV/{\AA}$^2$. Therefore, it should not be stable at least at low
temperatures. Furthermore, we found the Ga dimer length to be 2.4
{\AA} and no local minimum for Cerd\a'a's unusually large dimer
length.

The rehybridization of the $sp^3$ orbitals located at the threefold
coordinated Ga-atoms drives the relaxation towards a preferentially
flat Ga-bond configuration.  On the Ga terminated $\beta2 (4 \times
2)$ structure this leads to a decreased spacing between the Ga top
layer and the neighboring As layer which amounts to roughly half of
the bulk interlayer spacing. Also on the $\alpha (2 \times 4)$ and
$\beta2(2 \times 4)$ surfaces the threefold coordinated Ga atoms which
bond to As relax towards the plane of their neighboring As atoms.
Together with a slight upward shift of the top layer As atoms this
leads to a steepening\cite{ohno:93} of the As dimer block.  The change
of the angle between the bonds of the threefold coordinated As atoms
is less pronounced. However, the trend is obvious: except for the $c(4
\times 4)$ structure, we find the As bond-angles to be always smaller
than 109.5$^\circ$, which is the angle of the ideal tetrahedral
coordination. The As bonds on the $c(4 \times 4)$ surface behave
differently from those on the other three surfaces because the
top-layer As atoms are bonded to a second layer which consist of As
instead of Ga. A decrease of the angle between the bonds of all
threefold coordinated As atoms would require a change in the As-As
bond lengths, which probably costs more energy than would be gained
from rehybridization.

\subsection{(111) Surface}

The polar (111) orientation of GaAs has been studied within
density-functional theory by Kaxiras {\em et al}.\
\cite{kaxiras:86,kaxiras:86a,kaxiras:87}, who computed surface
energies relative to the ideal unreconstructed surface for various
atomic geometries. They found that under As-rich conditions an As
trimer geometry yields the lowest surface energy, whereas a Ga vacancy
reconstruction is preferred under Ga-rich conditions.  Haberern and
Pashley \cite{haberern:90} and Thornton {\em et al}.\
\cite{thornton:95} confirmed this experimentally. Haberern and Pashley
interpreted their STM images to show an array of Ga vacancies with a
(2$\times$2) periodicity. Thornton {\em et al}.\ observed both the As
triangle model and the Ga vacancy model in STM. Here we concentrate on
the following reconstructions of the Ga terminated (111) surface: the
As adatom, the As trimer, the Ga vacancy model, and, for comparison
but not as a reference system as in previous work, the truncated-bulk
geometry.

\begin{figure}[tb]
  \epsfxsize=6.9cm
  \hspace*{0.9cm}\epsfbox{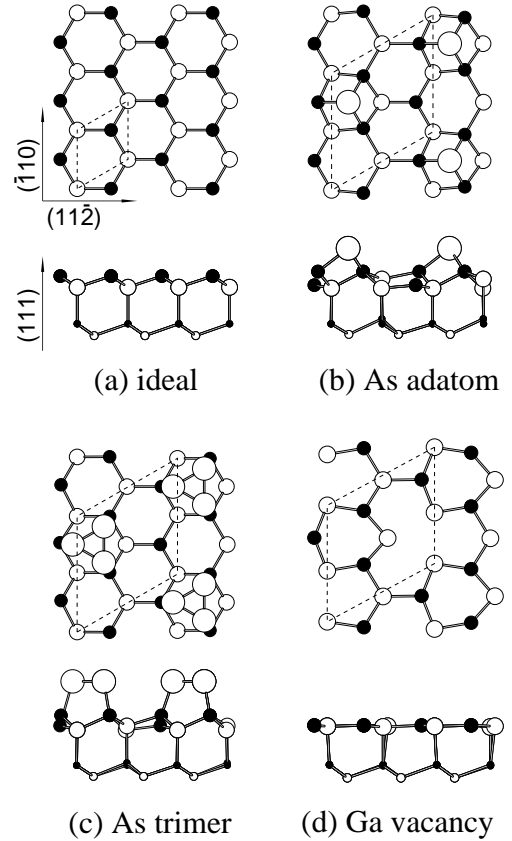}
  \caption{Atomic structures of the GaAs (111) reconstructions.}
  \label{111a_geo}
\end{figure}

The ideal (111) surface (see Fig.\ \ref{111a_geo}(a)) has 
a stoichiometry $\Delta N = -\frac{1}{4}/(1 \times 1)$. 
It does not fulfill the electron counting criterion.  Each Ga
dangling-bond is filled with 3/4 of an electron and therefore the
ideal surface has to be metallic. To create a neutral semiconducting
surface, following the electron counting criterion one can either add
an As surface atom to, or remove a Ga surface atom from, every $(2
\times 2)$ surface unit cell.  Therefore we consider three different
$(2 \times 2)$ reconstructions.  First of all, the As adatom model is
shown in Fig.\ \ref{111a_geo}(b).  This reconstruction is
stoichiometric. The As adatom binds to the Ga surface atoms.  It
exhibits a nearly orthogonal bond configuration, while the Ga atom
with the empty dangling bond relaxes towards the plane of the As
atoms. Secondly, we consider the As trimer model shown in Fig.\
\ref{111a_geo}(c). This model has a stoichiometry $\Delta N =
\frac{1}{2} /(1 \times 1)$, it also fulfills the electron counting
criterion and it is semiconducting. The three extra As atoms form a
trimer with each As atom binding to one Ga atom. The dangling-bonds of
the As atoms are completely filled and the dangling-bond of the the Ga
atom which is not bonded to As trimer atoms is completely empty.  This
Ga atom relaxes into the plane of the As atoms of the layer
below. Finally, we calculated the Ga vacancy model (see Fig.\
\ref{111a_geo}(d)). The removal of one Ga surface atom causes the
surface to be stoichiometric. The Ga surface atoms have completely
empty dangling-bonds and relax into the plane of the As atoms.  The
three As atoms surrounding the vacancy have completely filled
dangling-bonds.

\begin{figure}[tb]
  \epsfxsize=8.7cm
  \epsfbox{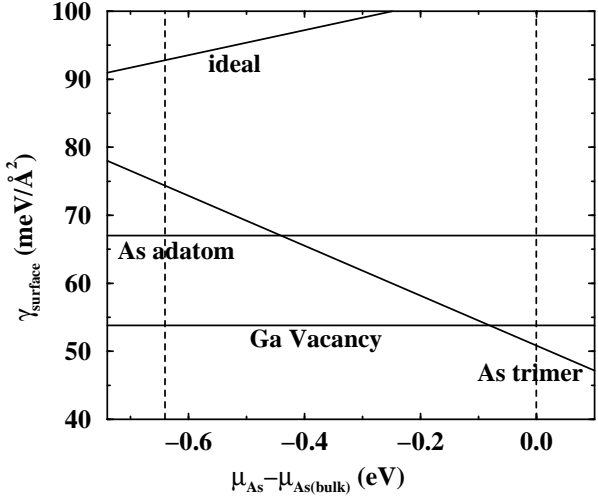}
  \caption{Surface energy of the different GaAs (111)
    reconstructions in meV/{\AA}$^2$ plotted versus the 
    difference of the chemical potential of As and As bulk.}
  \label{111a}
\end{figure}

We used the same super cell for the calculations of the (111) and the
(\=1\=1\=1) surfaces. Only the bulk-truncated surface was calculated
within a $(1 \times 1)$ surface unit cell, else always a $(2 \times
2)$ unit cell was used. The slab consisted of five (111) double
layers. The vacuum region had a thickness equivalent to four (111)
double layers. The whole Brillouin zone of the $(2 \times 2)$ surface
unit cell was sampled with 16 special {\bf k}-points, corresponding to
64 {\bf k}-points in the Brillouin zone of the $(1 \times 1)$ cell.
Absolute surface energies of the (111) reconstructions were determined
using the energy density formalism. The results are shown in Fig.\
\ref{111a}. 
The Ga vacancy model is the most favorable reconstruction for a large
range of the chemical potential from a Ga-rich to an As-rich
environment. Only in very As-rich environments the As trimer model has
a lower energy. The Ga vacancy model has a surface energy of 54
meV/{\AA}$^2$, whereas, the As trimer model has a surface energy of 51
meV/{\AA}$^2$ in As-rich environment at $\mu_{\rm As} = \mu_{\rm As
(bulk)}$. The Ga vacancy reconstruction was observed experimentally by
Haberern and Pashley \cite{haberern:90} and Tong {\em et al}.
\cite{tong:84} Thornton {\em et al}.\ additionally observed the As
trimer reconstruction.

Two other groups have performed similar {\em ab initio} calculations.
Using their energy density formalism, Chetty and
Martin\cite{chetty:92a} derived a value of 131 meV/{\AA}$^2$ for the
surface energy of the ideal (111) surface in a Ga-rich environment,
which is much larger than our value of 93 meV/{\AA}$^2$. Secondly, we
can compare our results to the relative surface energies of Kaxiras
{\em et al}.\cite{kaxiras:86a,kaxiras:87a} They arrived at the same
qualitative conclusions. However, quantitatively their relative
surface energies are not easily comparable to ours because they used
As$_4$ gas to define the As-rich environment.  Therefore they obtained
a larger interval for the chemical potential. We derive for the
surface energy difference of the As adatom and Ga vacancy structure a
value of 13 meV/{\AA}$^2$, whereas Kaxiras {\em et al}.\ calculate a
much larger difference of 47 meV/{\AA}$^2$.  Using their own result
for the ideal surface, Chetty and Martin transformed the relative
surface energies of Kaxiras {\em et al}.\ to absolute surface
energies. In comparison to our results, all these surface energies
contain the same shift towards higher energy as the ideal surface
mentioned above. We will discuss this difference below and explain, 
why we believe our results to be accurate.
 
Tong {\em et al}.\ \cite{tong:84} performed a LEED analysis for the
geometry of the Ga vacancy reconstruction. Their geometry data compare
very well with the theoretical data of Chadi \cite{chadi:84}, Kaxiras
{\em et al}.\ \cite{kaxiras:86a} and ours. For the Ga vacancy
reconstruction we find an average bond angle of the $sp^2$-bonded Ga
surface atom of 119.8$^\circ$ in agreement with Tong {\em et al}.  The
bond angles of the $p^3$-bonded As atom of 87.0$^\circ$ and
100.6$^\circ$ average to 91.5 $^\circ$ which again compare very well
with the value of 92.9$^\circ$ by Tong {\em et al}. The bonds of the
$p^3$-bonded As atom are strained by -1.6\% and 2.6\% with respect to
the GaAs bulk bonds. Tong {\em et al}.\ measured a value -1.3\% and
1.9\%, respectively.

Furthermore, for the As trimer reconstruction we compare our geometry
data to theoretical data of Kaxiras {\em et al}.  \cite{kaxiras:86a}
The threefold-coordinated As adatoms form bond angles to the
neighboring As adatoms of 60$^\circ$ due to symmetry reasons. The bond
angle of the As adatom to the next Ga atom is 106.2$^\circ$. Therefore
we get an average bond angle of 90.8$^\circ$ which is in good
agreement with the 91.7$^\circ$ of Kaxiras {\em et al}. The surface
Ga-As bonds are strained by 1.4 \%, whereas Kaxiras {\em et al}.\ find
the same bond length as in the bulk. The As-As bonds have a bond
length of 2.44 {\AA}, 2.4 \% shorter than that in As bulk. The Ga
surface atom which is not bond to an As adatom relaxes into the plane
of the As atoms with a bond angle of 118.4$^\circ$ and a bond length
which is 2.6 \% shorter than in GaAs bulk. These values are slightly
larger than the 114.7$^\circ$ and 1.0 \% reported by Kaxiras {\em et
al}.

\subsection{(\=1\=1\=1) Surface}

The polar GaAs (\=1\=1\=1) surface differs from the (111) surface, as
the bulk-truncated (\=1\=1\=1) surface is terminated by As atoms,
while the (111) surface is Ga terminated. At first sight the
(\=1\=1\=1) surfaces might seem to be still analogous to the (111)
surfaces, only that the Ga and As atoms have to be exchanged.
However, this analogy is not useful, because As and Ga have different
electronic properties, and therefore the (111) and (\=1\=1\=1)
surfaces do not exhibit equivalent reconstructions. Stoichiometric
(\=1\=1\=1) surfaces are gained by adding a Ga atom per $(2 \times 2)$
surface unit cell to the bulk-truncated surface or by removing an As
surface atom.

Kaxiras {\em et al}.\ \cite{kaxiras:87a} calculated the relative
surface energy for various $(2 \times 2)$ reconstructions. Biegelsen
{\em et al}.\ \cite{biegelsen:90} studied the (111) surface both
experimentally and theoretically. Using STM they observed an As trimer
$(2 \times 2)$ reconstruction for As-rich environments. A $(\sqrt{19}
\times \sqrt{19})$ reconstruction which is dominated by two-layer
hexagonal rings was identified for Ga-rich environments.

\begin{figure}[tb]
  \epsfxsize=6.9cm
  \hspace*{0.9cm}\epsfbox{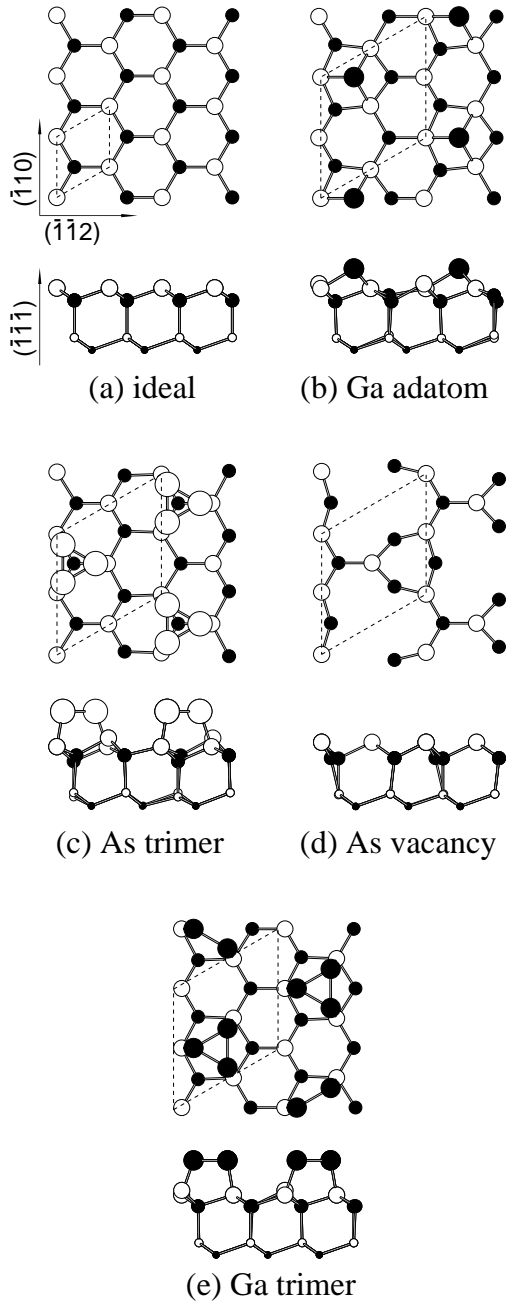}
  \caption{Atomic structures of the GaAs (\=1\=1\=1) reconstructions.}
  \label{111b_geo}
\end{figure}

Due to the large unit cell the $(\sqrt{19} \times \sqrt{19})$
reconstruction is computationally quite expensive, and in this work we
thus only consider $(2 \times 2)$ reconstructions.  First of all, for
comparison, we calculate the surface energy of the ideal (i.e.,
relaxed bulk-truncated) surface shown in Fig.\ \ref{111b_geo}(a). 
This surface is not stoichiometric $(\Delta N = \frac{1}{4} /(1 \times
1))$.  The dangling-bond of each As surface atom is filled with 5/4 of
an electron. Therefore the surface is metallic. Secondly, the Ga
adatom model shown in Fig.\ \ref{111b_geo}(b) was considered. Through
adding of an additional Ga surface atom the surface has become
stoichiometric and semiconducting. The dangling-bond of the Ga adatom
is completely empty, whereas the dangling-bond of the As atom which is
not bond to the Ga adatom is completely filled. Furthermore, we also
consider an As trimer model (see Fig.\ \ref{111b_geo}(c)). In contrast
to the (111) surface the As trimer is bond to As surface atoms. This
reconstruction has a stoichiometry of $\Delta N = 1$ per $(1 \times
1)$ surface unit cell.  Each As surface atom has a completely filled
dangling-bond.  Therefore, the surface is semiconducting.
Furthermore, we calculate the surface energy for the As vacancy model
which is shown in Fig.\ \ref{111b_geo}(d). The removal of the As
surface atom causes the surface to be stoichiometric. The three
neighboring Ga atoms have completely empty dangling-bonds. The surface
fulfills the electron counting criterion and is
semiconducting. Finally, we calculate the Ga trimer model (see Fig.\
\ref{111b_geo}(e)) to compare with Kaxiras {\em et al}.\
\cite{kaxiras:87a} and Northrup {\em et al}.  \cite{biegelsen:90} This
surface model has a stoichiometry of $\Delta N = -1/2$ per $(1 \times
1)$ surface unit cell and also fulfills the electron counting
criterion. However, it is metallic for the same reason as the Ga
terminated (110) surface.

\begin{figure}[tb]
  \epsfxsize=8.7cm
  \epsfbox{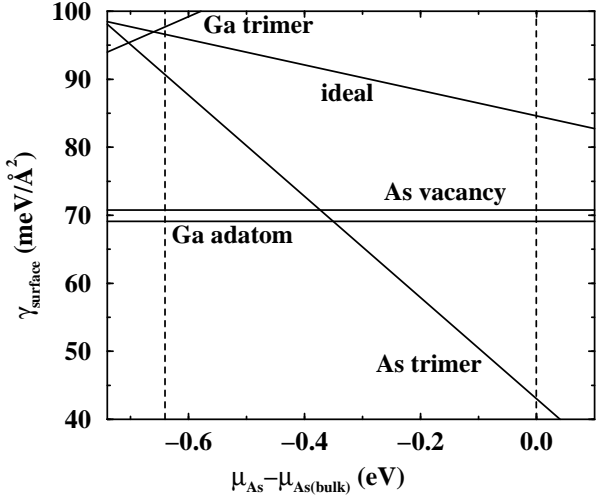}
  \caption{Surface energy of the different GaAs (\=1\=1\=1)
    reconstructions in meV/{\AA}$^2$ plotted versus the 
    difference of the chemical potential of As and As bulk.}
  \label{111b}
\end{figure}

The calculations for the (\=1\=1\=1) surface were carried out with the
same parameters and supercell as those for the (111) surface outlined
in the previous section. The results are shown in
Fig.\ \ref{111b}.
For As-rich environments we find that the As trimer model is the most
favorable reconstruction, as observed experimentally by STM and
confirmed by previous {\it ab initio} calculations.\cite{biegelsen:90}
This reconstruction has a very low surface energy of 43
meV/{\AA}$^2$. In a Ga-rich environment the Ga adatom reconstruction
has the lowest energy (69 meV/{\AA}$^2$) among all the structures we
calculated. The $(\sqrt{19} \times \sqrt{19})$ reconstruction found
experimentally was not included in our approach. However, as suggested
by Biegelsen {\em et al}.\ our present data can be used to restrict
the range of possible values for the surface energy of the $(\sqrt{19}
\times \sqrt{19})$ reconstruction consistent with observation: It has
to be smaller than the surface-energy of the Ga adatom model on the
one hand, and it has to be larger than the minimum energy of the
As-trimer surface (plus a small correction of $-3$ meV/\AA$^2$ to
account for the non-stoichiometry of the $\sqrt{19}\times\sqrt{19}$
reconstruction) on the other hand. Therefore, we conclude that energy
of the $(\sqrt{19} \times \sqrt{19})$ reconstruction is in the range
between 40 and 69 meV/{\AA}$^2$. Considering also the energetical
competition with facets of other orientations, even a slightly more
stringent condition can be deduced: For the (\=1\=1\=1) $(\sqrt{19}
\times \sqrt{19})$ surface in a Ga-rich environment to be stable
against faceting into \{110\} surfaces, its surface energy has to be
less than 63 meV/{\AA}$^2$.

In comparison to the relative surface energies calculated by Kaxiras
{\em et al}. \cite{kaxiras:87a} our energy difference between the As
vacancy and the Ga adatom structure of 2 meV/{\AA}$^2$ is only
slightly smaller than their value of 6 meV/{\AA}$^2$. However, they
state that for the Ga-rich environment the Ga trimer structure is 24
meV/{\AA}$^2$ more favorable than the Ga adatom structure.  In
contrast, we agree with Northrup {\em et al}.\ \cite{biegelsen:90}
that the Ga trimer is energetically quite unfavorable. It has a 29
meV/{\AA}$^2$ higher surface energy than the Ga adatom. Also the other
relative surface energies compare quite well with the already
mentioned calculations of Northrup {\em et al}. \cite{biegelsen:90}
although they derived a larger heat of formation (0.92 eV as opposed
to our value of 0.64 eV). Relative to the Ga adatom our surface
energies of the As trimer are about 10 meV/{\AA}$^2$ larger than
theirs. Also, they find a slightly larger energetic separation between
the As vacancy and Ga adatom structures. Their value for this energy
difference is 6 meV/{\AA}$^2$, whereas our value is 2
meV/{\AA}$^2$. However, these differences are small and do not affect
the physical conclusions. Chetty and Martin derived the absolute
surface energies using their result for the ideal (\=1\=1\=1) surface
and the relative surface energies of Kaxiras {\em et al}.\ and
Northrup {\em et al}. In contrast to the (111) their value of 69
meV/{\AA}$^2$ for the ideal (\=1\=1\=1) surface in the Ga-rich
environment is much smaller than ours of 97 meV/{\AA}$^2$. Therefore,
this time in comparison to our data the results all contain the same
shift to lower surface energies as the ideal surface. However, the sum
of the (111) and (\=1\=1\=1) surface energies from Chetty and Martin
is close to ours.  Therefore it is the splitting of the slab total
energy into contributions from the (111) and the (\=1\=1\=1) side that
comes out differently. In our calculations both sides are
energetically similar which seems to be plausible in view of the fact
that the flat (i.e., not faceted) surfaces have been observed
experimentally.

With respect to the calculated geometry we find that the As-As bond
length in the trimer is 2.46 {\AA}, 1.6 \% shorter than in bulk As.
The As trimer atoms each bind to an As atom 2.30 {\AA} beneath the As
trimer plane in agreement with Northrup {\em et
al}. \cite{biegelsen:90} The remaining As atom which is not bond to
the trimer relaxes outwards and is 1.74 {\AA} below the trimer plane.
This compares reasonably well with the slightly larger value of 1.89
{\AA} by Northrup {\em et al}. For the two Ga surface models the
separation of the adatom or trimer plane and the closest As (rest atom)
plane amounts to 0.98 {\AA} for the Ga adatom model, and 1.98 {\AA} for the
Ga trimer model. Northrup {\em et al}.\ derived values of 0.98
{\AA} and 1.90 {\AA}.

\subsection{Equilibrium Crystal Shape (ECS)}

As opposed to liquids, crystals have non-trivial equilibrium shapes
because the surface energy $\gamma(\hat{\bf m})$ depends on the
orientation $\hat{\bf m}$ of the surface relative to the
crystallographic axes of the bulk. Once $\gamma(\hat{\bf m})$ is
known, the ECS is determined by the Wulff construction,
\cite{wulff:01,wortis:88} which is equivalent to solving
\begin{equation}
  r(\hat{\bf h}) = \min_{\hat{\bf m}} 
  \left(\frac{\gamma({\hat{\bf m}})} {\hat{\bf m} \cdot \hat{\bf h}} \right).
\end{equation}
Here $r(\hat{\bf h})$ denotes the radius of the crystal shape in the
direction $\hat{\bf h}$. When the surface energy $\gamma(\hat{\bf m})$
is drawn as polar plot, the ECS is given by the interior envelope of
the family of planes perpendicular to $\hat{\bf m}$ passing through
the ends of the vectors $\gamma(\hat{\bf m})\,\hat{\bf m}$. Under the
assumption that only the (110), (100), (111), and (\=1\=1\=1) facets
exist, we construct the ECS from the calculated surface energies of
these facets. Thus there may exist additional thermodynamically stable
facets that are missing on our ECS. To be sure to construct the
complete shape one would have to calculate the surface energy for
every orientation. However, from experiments it is known that the low
Miller-indices surface orientations we consider are likely to be the
energetically most favorable ones.

\begin{figure}[tb]
  \epsfxsize=8.7cm
  \epsfbox{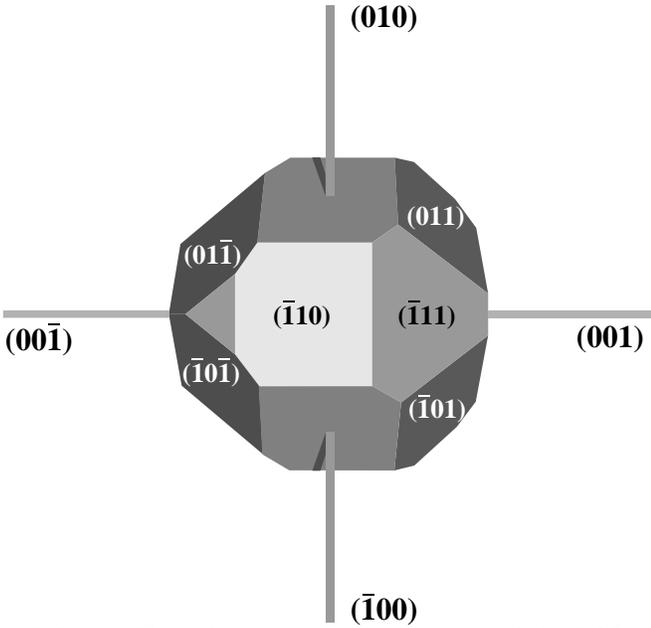}
  \caption{Three-dimensional representation of the ECS of GaAs in an 
    As-rich environment, constructed from the surface energies of the 
    (110), (100), (111), and (\=1\=1\=1) facets. The (\=100), (010), and 
    (001) axes are drawn for convenience.}
  \label{3d_ecs}
\end{figure}

\begin{figure}[tb]
  \epsfxsize=7cm
  \hspace*{0.85cm}\epsfbox{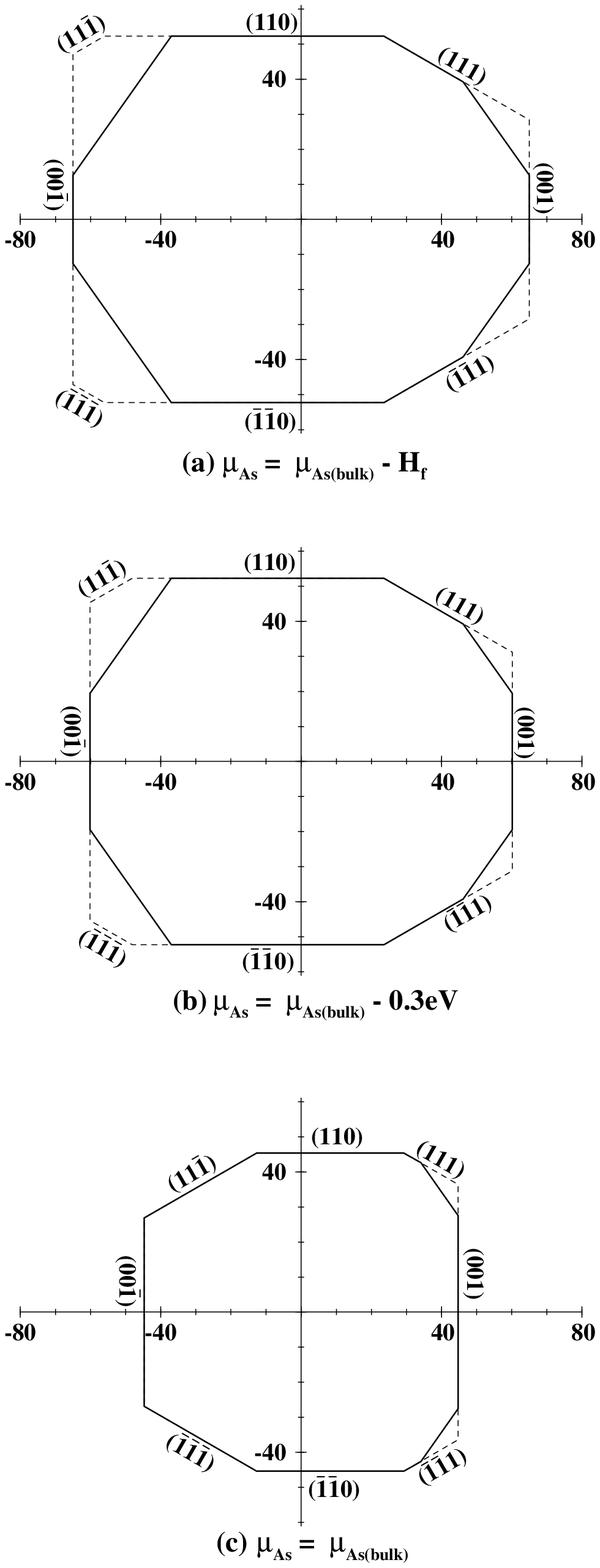}
  \caption{Cross section of the ECS of GaAs for three different chemical 
    potentials $\mu_{\rm As}$. The dashed line denotes the equilibrium 
    shape of an infinitely long cylindrical crystal, derived from a 
    two-dimensional Wulff-construction. The (\=1\=11) orientation is 
    equivalent to the (111) and the (11\=1) to the (\=1\=1\=1).}
  \label{ecs}
\end{figure}

As the GaAs surface energies depend on the chemical environment, the
ECS becomes a function of the chemical potential. In Fig.\ \ref{3d_ecs} 
the ECS is shown for an As-rich environment and zero temperature. 
The different facets have been marked in the Figure and the ECS reflects
the symmetry of bulk GaAs. To investigate the dependence of the ECS on the 
chemical potential we will focus on the cross-section of the ECS with a 
(1\=10) plane through the origin. This cross-section includes the complete 
information from all four calculated surfaces, because they all possess 
surface normals within this plane. The ECS is shown for three different 
chemical environments in
Fig.\ \ref{ecs}. 
Note that in a Ga-rich environment the (\=1\=1\=1)$(\sqrt{19} \times
\sqrt{19})$ reconstruction would be energetically more favorable than
the (\=1\=1\=1)(2$\times$2) Ga-adatom reconstruction used for the
construction of the ECS at this chemical potential, i.e., the
experimental (\=1\=1\=1) facet appears somewhat closer to the
origin. For an As-rich environment we find that all four considered
surface orientations exist in thermodynamic equilibrium. Furthermore,
the (111) surface exists within the full range of accessible chemical
potentials.  This is in contrast to the result Chetty and Martin
\cite{chetty:92a} derived from the work of Kaxiras {\em et al}.\
\cite{kaxiras:87}: They stated that the (111) surface has a high
energy and thus it should not exist as a thermodynamic equilibrium
facet. However, experimental work of Weiss {\em et al}.\
\cite{weiss:89} using a cylindrical shaped sample indicates that
between the (110) and (111) orientation all surfaces facet into (110)
and (111) orientations. The $(2 \times 2)$ superstructure of the (111)
surface has been observed on these faceted surfaces. If the (111)
orientation of GaAs were instable, the appearance of facets other than
(111) on the cylindrical crystal were to be expected.

In Fig.\ \ref{ecs} one can see that the ECS becomes smaller for
As-rich environments. The As terminated reconstructions have surface
energies about 20 \% smaller than those found in Ga-rich environments,
which are mostly stoichiometric like the Ga vacancy. In contrast to
the surface reconstructions found for As-rich environments no similar
Ga terminated reconstructions are observed. Another remarkable feature
of the ECS is that the surface energies do not vary very much with the
orientation. For Ga-rich environments they vary by about $\pm 10 \%$,
whereas for As-rich environments they vary only by $\pm 5 \%$.

Our calculated ECS imposes restrictions on the surface energies of
other surface orientations: When it has been proven experimentally
that a facet exists in thermodynamic equilibrium, one can derive a
lower and an upper limit for its surface energy. The limits are given
by the surface energy of the neighboring facets on our ECS together
with appropriate geometry factors. They follow from the conditions
that (a) the surface energy has to be sufficiently small, so that the
surface does not facet into \{110\}, \{100\}, \{111\}, and
\{\=1\=1\=1\} orientations, and (b) that the surface energy is not so
small that neighboring facets are cut off by this plane and thus
vanish from the ECS.  In a similar way the Wulff construction yields a
lower limit for the surface energy of any facet that does not exist in
thermodynamic equilibrium.

Recently the shape of large three-dimensional InAs islands (diameter
$\sim$ 2000 \AA) grown by MOVPE on a GaAs(100) substrate has been
observed by E. Steimetz {\em et al.} \cite{steimetz:96} These islands
are presumably relaxed, the misfit of the lattice constants being
compensated by a dislocation network at the InAs-GaAs interface.  Thus
the facets displayed on these islands should be identical to the
facets on the ECS of InAs. In fact, the observed shapes are compatible
with an ECS like that of GaAs shown in Fig.\ \ref{3d_ecs}, with
\{110\}, \{100\}, \{111\}, and \{\=1\=1\=1\} facets being clearly
discernible. Due to the similarity between InAs and GaAs we take this
as another confirmation of our results as opposed to those of Chetty
and Martin. \cite{chetty:92a}

\section{Summary and Conclusion}

The GaAs surface energies of different orientations have been
calculated consistently with one and the same parameters and
pseudo-potentials. The surface energies of the (110), (100), (111) and
(\=1\=1\=1) surfaces are given in dependence of the chemical
potentials.

For the (111) and (\=1\=1\=1) surfaces we find a large difference to
previous results of Chetty and Martin. \cite{chetty:92a} They derived
a difference of about 62 meV/{\AA}$^2$ between the surface energies of
the ideal (111) and (\=1\=1\=1) surfaces, whereas we calculate a
difference of about $-4$ meV/{\AA}$^2$. Consequently the absolute
surface energies calculated by Chetty and Martin using data of Kaxiras
{\em et al}.\ \cite{kaxiras:87,kaxiras:87a} and Northrup {\em et al}.\
\cite{biegelsen:90} contain the above difference of 66
meV/{\AA}$^2$. This is due to a different splitting of the slab energy
into contributions from the (111) and (\=1\=1\=1) surfaces, as Chetty
and Martin's and our sum of the (111) and (\=1\=1\=1) surface energies
are essentially equal. Obtaining high surface energies for the (111)
surfaces Chetty and Martin have to conclude that the (111) facet
should be unfavorable and not exist in thermodynamic equilibrium. In
contrast our surface energies for the (111) surface are lower and
therefore we conclude that it exists in thermodynamic equilibrium
which appears to be in agreement with experimental observations.

As already stated by Chetty and Martin \cite{chetty:92a} there are
significant differences between the results of Kaxiras {\em et al}.\
and Northrup {\em et al}.\ for the (\=1\=1\=1) surface: Kaxiras {\em
et al}.\ find the Ga trimer structure to be energetically favorable in
Ga-rich environments, whereas we agree with Northrup {\em et al}.\ and
find it energetically unfavorable. This is also confirmed by
experiment.

Having calculated the absolute surface energies for different
orientations we are in the position to construct the ECS of GaAs. We
have to keep in mind, however, that it is implicitly assumed that only
the (110), (100), (111) and (\=1\=1\=1) surfaces exist in
equilibrium. For a more refined discussion of faceting further
calculations also for higher-index surfaces would have to be
performed. From our ECS we conclude that in As-rich environment all
four orientations exist in thermodynamic equilibrium. For a given
chemical potential the variation of the surface energy with
orientation is small and less than $\pm 10 \%$. Our ECS of GaAs gives
indication for the ECS of InAs or other III-V semiconductors which
show similar surface reconstructions.

\section{Acknowledgments}

We thank E. Steimetz for helpful discussion and a copy of 
Ref.\ \onlinecite{steimetz:96} prior to publication.
This work was supported in part by the Sfb 296 of the Deutsche 
Forschungsgemeinschaft.

\end{document}